\documentclass[aps,pre,%
preprint,%
amssymb,amsmath,%
nopreprintnumbers,%
showpacs,%
showkeys,%
fleqn,%
eqsecnum,%
byrevtex]{revtex4}
\usepackage{calc}
\usepackage[final]{graphicx}
\begin{document}
\title{The new screening characteristics of strongly non-ideal and dusty plasmas.
Part 2: Two-Component Systems}
\author{A. A. Mihajlov$^{1}$}
\email{mihajlov@phy.bg.ac.rs}
\author{Y. Vitel$^{2}$}
\email{yv@ccr.jussieu.fr}
\author{Lj. M. Ignjatovi\'c$^{1}$}
\email{ljuba@phy.bg.ac.rs}

\affiliation{$^{1}$Institute of Physics, P.O. Box 57, 11080 Zemun,
Belgrade, Serbia} \affiliation{$^{2}$Laboratoire des Plasmas Denses,
Universite P. et M. Curie, 3 rue Galilee, Paris, 94200 Ivry sur
Seine, France}
\begin{abstract}
A new model method for describing of the electrostatic screening in
two-component systems (electron-ion plasmas, dusty plasmas,
electrolytes, etc) is developed. The method is applicable to the
systems of higher non-ideality degree. The expressions for all the
screening parameters introduced in the previous paper (Part 1) of
this work, as well as for an additional parameter characteristic for
multi-component systems, are obtained. All these parameters are
presented in a simple analytic form suitable for operative
laboratory usage, especially for theoretical interpretation of
experimental data.
\end{abstract}
 \maketitle
\section{Introduction}
\label{sec:intro}

In the previous paper \cite{mih08}, here Part 1, the aims of this
research were already described, as well as the stimuli for its
starting. Accordingly to these aims the new model method of
describing of the electrostatic screening in electron-ion plasmas
and other two-component systems (e.g. dusty plasmas and some
electrolytes) which relies on the basic model (a1) - (a3) is
presented in this part. This method is free of non-physical
properties of Debye-H\"{u}ckel's (DH) method and posses the positive
features (b1) and (b2) described in Section 1 of Part 1, together
with the basic model.

The material presented in this paper is distributed in the four
following Sections and four Appendixes. Section~\ref{sec:tas}
contains: the screening model; the critical analysis of DH method in
the case of two-component system; stating the tasks precisely. In
Section~\ref{sec:mp} and Section~\ref{sec:mdie} the developed method
and obtained solutions for two-component systems are presented. The
Section~\ref{sec:pso} contains obtained results and discussion.

\section{Theory assumptions}
\label{sec:tas}

\subsection{Screening model}
\label{sec:cas}

A stationary homogeneous two-component system $S_{in}$ is taken here
as the initial model of some real physical objects. We will assume
that $S_{in}$ is constituted by a mix of two gases: one of positive charged
ions (of only one kind), and other of electrons. It is assumed that
these gases there are in the equilibrium states with temperatures
$T_{i}$ and $T_{e} \ge T_{i}$, and mean local particle density $N_{e}$ and
$N_{i}$. All the particles are treated as point objects with the
charge $Z_{e}e$ in the case of electron, and $Z_{i}e$ in the case of
ion, where $Z_{e}=-1$, $Z_{i}=1,2,...$, and $e$ is the modulus of
the electron charge. Let us note that in this paper the electron
charge will be also denoted by $-e$. It is understood that the
parameters $Z_{e,i}$ and $N_{e,i}$ satisfy the local
quasi-neutrality condition
\begin{equation}
\label{eq1}  Z_i e \cdot N_i - e \cdot N_e = 0,
\end{equation}
as well as that $N_{e}$ and $T_{e}$ allow the non-relativistic
treatment of the electron component.

In accordance with the properties (a1) and (a2), the screening of a
charged particles in the system $S_{in}$ will be modeled in the
corresponding accessory systems each of which differs from $S_{in}$
in that, besides the two described components, it also contains a
fixed probe particle with charge $Z_{p}e$ in the origin of the
chosen reference frame (point $O$). Here we will study two cases:
the ion case $(i)$, when $Z_p = Z_i$, and the electron case $(e)$,
when $Z_p = Z_{e} = - 1$, when the probe particle represents one of
the particles of the system $S_{in}$. In accordance with this, we
will denote here the corresponding accessory system with
$S_{a}^{(i)}$ or $S_{a}^{(e)}$. This system will be characterized
by: the ion and electron densities $n_i^{(i,e)} (r)$ and
$n_e^{(i,e)} (r)$, the mean local charge density
\begin{equation}
\label{eq2}  \rho ^{(i,e)}(r) = Z_i e \cdot n_i^{(i,e)} (r) - e
\cdot n_e^{(i,e)} (r),
\end{equation}
and the mean electrostatic potential $\Phi^{(i,e)}(r)$, where
$r=|\vec{r}|$ and $\vec{r}$ is the radius-vector of the observed
point. It is assumed the satisfying of the boundary conditions
\begin{equation}
\label{eq3}  \lim\limits_{r \to \infty } n_i^{(i,e)} (r) = N_i,
\qquad \lim\limits_{r \to \infty } n_e^{(i,e)} (r) = N_e ,
\end{equation}
and the condition of neutrality of the systems $S_p^{(i,e)} $ as
wholeness
\begin{equation}
\label{eq4}  Z_{i,e} e + \int\limits_0^\infty {\rho ^{(i,e)}(r)}
\cdot 4\pi r^2dr = 0.
\end{equation}
Then, we will take into account that the $\Phi ^{(i,e)}(r)$ and
$\rho^{(i,e)}(r)$ have to satisfy Poisson's equation
\begin{equation}
\label{eq5}  \nabla^{2} \Phi ^{(i,e)} = - 4\pi \left[ Z_{i,e}e\cdot
\delta(\vec r) + \rho ^{(i,e)}(r) \right],
\end{equation}
where $\delta(\vec r)$ is three-dimensional delta function
\cite{ich73}. From the same reason as in Part 1, this equation
applies in the whole region $r > 0$. It is assumed the satisfying
the boundary conditions
\begin{equation}
\label{eq6}  \lim\limits_{r \to \infty } \Phi ^{(i,e)}(r) = 0,
\end{equation}
\begin{equation}
\label{eq7}   \left| {\varphi ^{(i,e)}} \right| < \infty, \qquad
\varphi ^{(i,e)} \equiv \lim\limits_{r \to 0} [\Phi ^{(i,e)}(r) -
Z_{i,e}e/r].
\end{equation}
Since $\varphi ^{(i,e)}$ is the mean electrostatic potential in the
point $O$, the quantity
\begin{equation}
\label{eq8}  U^{(i,e)} = Z_{i,e} e \cdot \varphi ^{(i,e)}.
\end{equation}
is the potential energy $U^{(i,e)}$ of the probe particle. In an
usual way $U^{(i)}$ and $U^{(e)}$ are treated as approaches to the
mean potential energies of the ion and electron in the system
$S_{in}$.

In accordance with the properties (a2) and (a3) the conditions of
thermodynamical equilibrium of the ion component in the case $(i)$,
as well as the electron component in the case $(e)$, will be taken
in the form
\begin{equation}\label{eqmu1}
 \mu_{i,e}\left( {n_{i,e}^{(i,e)}(r),T_{i,e} } \right)+Z_{i,e} e
\cdot \Phi^{(i,e)}(r)=\mu_{i,e} \left({ N_{i,e} ,T_{i,e} }\right)
\end{equation}
where $\mu _i (n_i^{(i)} (r),T_i )$ and $\mu _e (n_e^{(e)} (r),T_e
)$ are the chemical potentials of the ideal ion and electron gases,
which can depend of the corresponding particle spins,
considered on the distance $r$ from the point $O$. On the base of
the considerations from Part 1 one should keep in mind that the
equations (\ref{eqmu1}) are applicable only in the regions
\begin{equation}\label{eqrrs}
 r \ge r_{s;i,e}, \qquad r_{s;i,e} \equiv \left(\frac{3}{4\pi
 N_{i,e}}\right)^{1/3},
\end{equation}
where $r_{s;i} $ and $r_{s;e} $ are the corresponding Wigner-Seitz's
radii. In the used procedure the equation (\ref{eqmu1}) is taken in
the linearized form
\begin{equation}
\begin{array}{c}
\label{eq9}  n_{i,e}^{(i,e)} (r) - N_{i,e} = {\displaystyle -
\frac{Z_{i,e} e}{\partial \mu _{i,e}/\partial N_{i,e}}\Phi
^{(i,e)}(r)}, \qquad \displaystyle{{\partial\mu_{i,e}}/{\partial
N_{i,e}}\equiv \left[\frac{\partial \mu_{i,e} (n,T_{i,e})}{\partial
n}\right]_{n = N_{i,e}}},
\end{array}
\end{equation}
but under the condition
\begin{equation}\label{eq13a}
\frac{|n_{i,e}^{(i,e)}(r)-N_{i,e}|}{N_{i,e}} \ll 1.
\end{equation}
It is important that the conditions (\ref{eqrrs}) and (\ref{eq13a})
are compatible in all considered cases.

Since we take the single-component systems considered in Part 1, as
a boundary case of two-component systems (when it is spread one of
their components), we will require that the ion density
$n_{i}^{(i)}$ in the case (i) and the electron density $n_{e}^{(e)}$
in the case (e) satisfy the equation
\begin{equation}
\label{eq13}  \int\limits_0^\infty {\left[ {N_{i,e} -
n_{i,e}^{(i,e)} (r)} \right] \cdot 4\pi r^2} dr = 1.
\end{equation}
which is analogous of the equation (21) from Part 1. From
Eqs.~(\ref{eq4}) and (\ref{eq13}) it follows that the electron
density $n_{e}^{(i)}$ in the case (i) and ion density $n_{i}^{(e)}$
in the case (e) have to satisfy another equation
\begin{equation}
\label{eq14}  \int\limits_0^\infty {\left[ {n_{e,i}^{(i,e)} (r) -
N_{e,i}} \right] \cdot 4\pi r^2} dr = 0.
\end{equation}
Let us emphasize that the relations (\ref{eq13}) and (\ref{eq14})
can be obtained on the base of the interpretation of the systems
$S_{a}^{(i,e)}$ which is given in Appendix~\ref{sec:appc}. In
further considerations is used the fact that simultaneous satisfying
of the conditions (\ref{eq13}) and (\ref{eq14}) automatically
provide the satisfying of neutrality condition (\ref{eq4}).

\subsection{The critical analysis of DH method} \label{sec:sdm}

The procedure of obtaining of DH solutions is described in
Appendix~\ref{sec:appa}. The figure~\ref{fig:nrhoD} shows the
behavior of the particle densities $n_{D;i}^{(i)}(r)$ and
$n_{D;e}^{(i)} (r)$, and charge density $\rho _D^{(i)} (r)$. This
figure illustrates that the behavior of DH densities of the free
particles with the same charge as the probe particle is
qualitatively same as in the case which is considered in the Part 1.
Consequently, the procedure of the elimination of the non-physical
properties of $n_{D;i}^{(i)}(r)$ in the case (i), and $n_{D;e}^{(i)}
(r)$ in the case (e) will be similar to the procedure which is
described in Part 1.

\begin{figure}[htbp]
\centerline{\includegraphics[width=\columnwidth,
height=0.75\columnwidth]{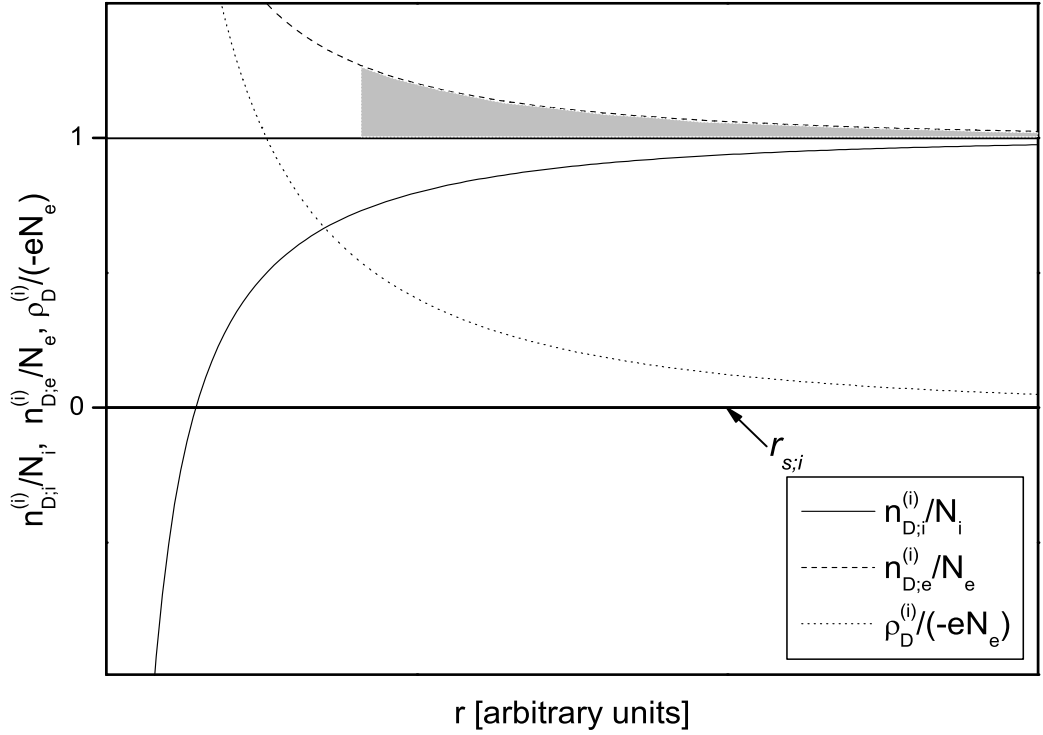}} \caption{The reduced DH
densities $n_{D;i}^{(i)} (r) / N_{i}$, $n_{D;e}^{(i)} (r) / N_{e}$
and $\rho _{D}^{(i)} (r) / (-eN_{e}) $ in the case $Z_{i}=1$,
$T_{e}=T_{i}$ and $\kappa_{D} r_{s;i} = 1$, where $\kappa_{D} $ is
Debye's screening constant given by (\ref{eqA1}). Shadowing
emphasize the non-physical deviation of $n_{D;e}^{(i)}$ from the
asymptotic value $N_{e}$.} \label{fig:nrhoD}
\end{figure}

The main disadvantage of DH method consists in the monotonous
increasing of DH densities of the free particles, which the charge
is opposite to the charge of the probe particle, with the decreasing
of $r$ in the whole region $r < \infty$. This fact is illustrated by
the behavior of $n_{D;e}^{(i)} (r)$ in Fig.~\ref{fig:nrhoD}. Because
of such a behavior DH solutions principally can not satisfy the
conditions (\ref{eq14}). Namely, from (\ref{eqA5}) it follows that
in DH case the left side of those conditions is not equal to $0$,
but it is proportional to $Z_{i}$ in the case $(i)$, and $(1/Z_{i})$
in the case $(e)$. The consequence of this fact is the principal
impossibility to treat the probe particle as a represent of a
particle in the system $S_{in}$. For an example, in the case of
completely classical electron-ion plasma with $Z_{i}=1$ and
$T_{i}=T_{e}$ from the non-satisfying of the conditions (\ref{eq14})
it follows that the mean number of electrons per ion should be
$3/2$, instead of $1$.

The described disadvantage is a consequence of two facts: that in DH
method the first step is determining of electrostatic potential
$\Phi _D^{(i;e)} (r)$ in the whole space, and the equations
(\ref{eq9}) are used together with the equations (\ref{eq10}) from
Appendix~\ref{sec:appa}. Because of that within DH method the
electron and ion components are "smeared" in the space
simultaneously and independently in both (i) and (e) cases.

\subsection{What one should do in order to avoid non-physicality of
DH method?}\label{sec:what}

In accordance with above mentioned, our main task is finding of such
a procedure within the basic model (see Section~1 in Part 1), which
would be alternative one to DH procedure. It assumes that sought
procedure has to provide the possibility of determining of the
solutions $n_{i}^{(i)}(r)$ and $n_{e}^{(i)} (r)$ in the same way as
it was described in Part 1, and the solutions $n_{e}^{(i)}(r)$ and
$n_{e}^{(i)} (r)$ without using the equations (\ref{eq10}) from
Appendix~\ref{sec:appa}. Also, the satisfying of the conditions
(\ref{eq13}) and (\ref{eq14}) is assumed.

\section{The presented method: the electron and ion densities}
\label{sec:mp}

\subsection{The case (i)}
\label{sec:mdmi}

\begin{figure}[htbp]
\centerline{\includegraphics[width=\columnwidth,
height=0.75\columnwidth]{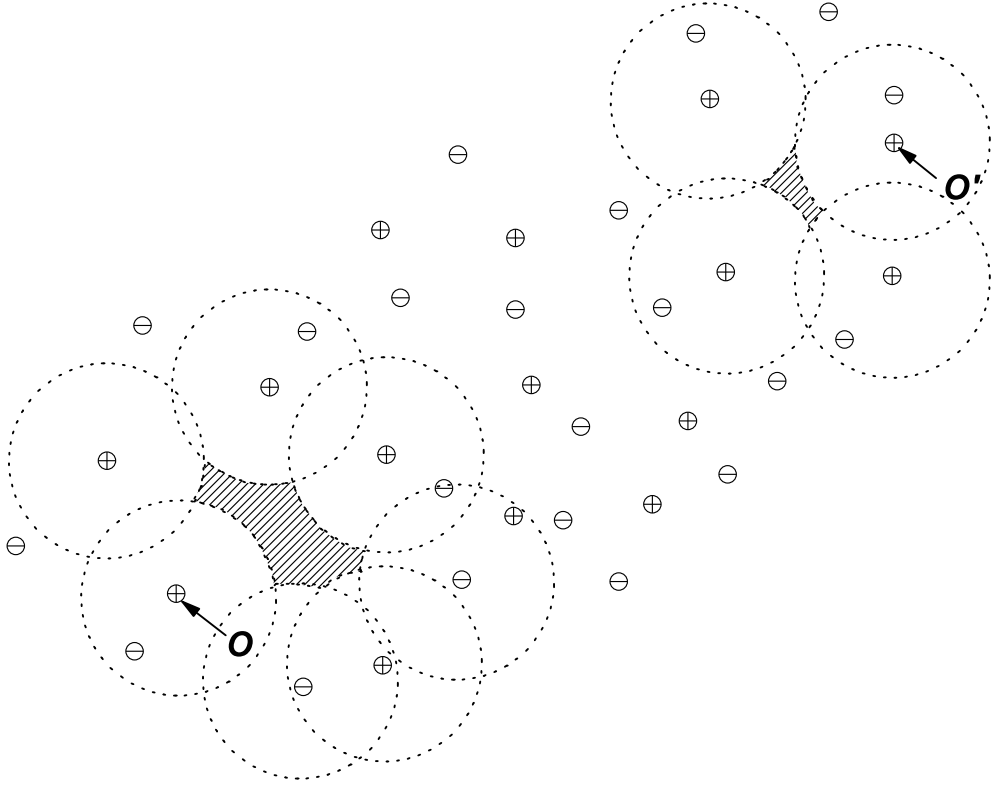}} \caption{The schematic
picture of ion-electron plasma. Shadowed areas represent the regions
of $\vec{r}$ without of probe particle's and all ion's self-shells.
With $O$ and $O'$ the places of the probe particle in the case $(i)$
and one of ions are denoted, and with the dotted lines probe
particle and ions self spheres are shown.} \label{fig:scheme}
\end{figure}

{\bf The solution $n_{i}^{(i)}(r)$.} In order to solve our task we
 start from such a picture of the system $S_{a}^{(i)}$ where
electrons are treated in the electrostatic field of the probe
particle and all ions (treated in a classic way) which are
distributed in discrete points, as it is illustrated by
Fig.~\ref{fig:scheme}. On the base of this picture, the procedure of
the expressing of $n_{e}^{(i)}(r)$ in the region $r > r_{s;i}$
trough $n_{i}^{(i)}(r)$ is developed in the first part of Appendix
\ref{sec:appb1}. The results of this procedure is the relation
(\ref{eq:22d}), which can be presented in the form
\begin{equation}
\label{eq22}  n_e^{(i)} (r) = N_{e}\cdot (1-\alpha) + \alpha \cdot
Z_i n_i^{(i)} (r), \qquad r_{s;i} < r < \infty,
\end{equation}
where the parameter $\alpha$ will be determined in further text.
This relation, together with Eq.~(\ref{eq2}), makes possible to
represent the charge density $\rho^{(i)}(r)$  in the form
\begin{equation}
\label{eq22a}  \rho^{(i)}(r) = Z_{i}e \cdot (1-\alpha)
\cdot[n_{i}^{(i)}(r) - N_{i}], \qquad r_{s;i} < r < \infty,
\end{equation}
and to use for determination of $n_i^{(i)} (r)$ the procedure which
is described in details in Sections~3 and 6 of Part 1. As first, by
means, Eqs.~(\ref{eq9}) and (\ref{eq22a}) it is obtained the
equation of Volterra's type, namely
\begin{equation}
\label{eq22b}  \rho^{(i)}(r) = \kappa_{i} ^2\int\limits_r^\infty
{\rho^{(i)} (r')\left( {\frac{1}{r} - \frac{1}{r'}} \right)}
r'^2dr',
\end{equation}
where
\begin{equation}
\label{eq29}  \kappa _i \equiv \frac{1}{r_{\kappa ;i} } = \kappa
_{0;i} \cdot \left( {1 - \alpha } \right)^{\textstyle{1 / 2}},
\qquad \kappa _{0;i} =\left[ \frac{4\pi (Z_{i} e)^2} {\partial \mu
_i/\partial N_{i} } \right]^{1 \over 2}.
\end{equation}
One can see that in the two-component case the ion screening
constant $\kappa _i $ and the corresponding characteristic length
$r_{\kappa ;i}$ depend on the parameter $(1 - \alpha)^{1 \over 2}$.

Accordingly to Part 1, the solution of the equation (\ref{eq22b}) in
the general case can be taken in the form: $\rho^{(i)}(r) = c_{i}
\cdot Z_{i} e \cdot \exp(-\kappa_{i} r) / r$. From here and
Eq.~(\ref{eq22b}) it follows that ion density $n_{i}^{(i)}(r)$ in
the region $r_{s;i} < r < \infty$ is given by the relation:
$n_{i}^{(i)}(r) = N_{e} - c_{i} \cdot \exp(-\kappa_{i} r) / r$.
Than, taking $c_{i} = r_{0;i} \cdot \exp(\kappa_{i} r_{0;i}) \cdot
N_{e}$ and applying to this relation the extrapolation procedure
from Part 1, we obtain the sought expression for $n_{i}^{(i)}(r)$ in
the whole space, namely
\begin{equation}
\label{eq34}  n_i^{(i)} (r) = \left\{ {{\begin{array}{*{20}c}
 \displaystyle{ {N_i - N_i r_{0;i} \cdot \exp (\kappa _i r_{0;i} ) \cdot
\frac{\exp ( - \kappa _i
r)}{r},}} \hfill & {r_{0;i} < r ,} \hfill \\
 {0,} \hfill & {r \le r_{0;i} ,} \hfill \\
\end{array} }} \right.
\end{equation}
where the new characteristics length $r_{0;i} $ has to be determined
from the condition (\ref{eq13}). Using the results of Part 1 we can
present the radius $r_{0;i}$ in two equivalent forms
\begin{equation}
\label{eq35}  r_{0;i} = \gamma_{s}(x_{i})\cdot r_{s;i}, \qquad
r_{0;i} = \gamma_{\kappa}(x_{i})\cdot r_{\kappa;i},
\end{equation}
\begin{equation}
\label{eq36}  x_{i} = \kappa_{i} r_{s;i} ,
\end{equation}
where the coefficients $\gamma_{s}(x)$ and $\gamma_{\kappa}(x)$ are
given by Eq. (28) from Part 1, and $r_{s;i}$ - by Eq.~(\ref{eqrrs}).

{\bf The solution $n_{e}^{(i)}(r)$.} In order to determine
$n_{e}^{(i)}(r)$ in the probe particle self-sphere ($0 < r \le
r_{s;i}$), we will take into account that in the system $S_{in}$ the
mean number of electrons in the sphere with volume $1/N_{i}$, which
is centered at some ion (the ion self-sphere), is larger than the
mean number of electrons in every fixed sphere with the same volume,
i.e. $N_{e}\cdot (1/N_{i}) = Z_{i}$, because of the additional
electrons whose presence is caused by presence of other ions. In
accordance with this we will find $n_{e}^{(i)}(r)$ in the probe
particle self-sphere in the system $S_{a}^{(i)}$ in the form
\begin{equation}\label{eqnes2}
 n_{e}^{(i)}(r) =  n_{e;s}(r) + n_{e;ion}(r), \quad 0 < r < r_{s;i},
\end{equation}
where the member $n_{e;s}(r)$ satisfies the condition
\begin{equation}
\label{eq24}  \int\limits_0^{r_{s;i} } {n_{e;s} (r) \cdot 4\pi r^2dr
= Z_{i} },
\end{equation}
considering that this member describes the distribution of $Z_{i}$
electrons which there are in the probe particle self-sphere
independently of the presence of ions, while the member
$n_{e;ion}(r)$ describes the distribution of the mentioned
additional electrons. Let us emphasize that such a treatment of the
difference $[n_{e}^{(i)}(r) -  n_{e;s}(r)] \equiv n_{e;ion}(r)$ is
caused by the procedure of the obtaining the member $n_{e;s}(r)$
which is described in the second part of Appendix~(\ref{sec:appb1}).

In order to establish the connections between the members in
Eqs.~(\ref{eq22}) and (\ref{eqnes2}), we will assume that the form
of the equation (\ref{eq22}), which transforms at $r=\infty$  to
equality $N_{e} = N_{e}(1-\alpha) + N_{e}\alpha$, reflects the
existing in the system $S_{in}$ of the spatial correlation between
electrons and ions. It means exceptionally following: in every fixed
volume $V$ in average there are $V \cdot N_{e}(1-\alpha)$ electrons
independently of presence of ions, and  $V \cdot N_{e}\alpha$
electrons whose presence is caused by the presence of ions. The fact
that the presence of ions in some volume causes the presence of
additional electrons itself has already been used above, while here
it is assumed that just the parameter $\alpha$ in Eq.~(\ref{eq22})
represents the quantitative characteristic of the mentioned
electron-ion correlation in the system $S_{a}^{(i)}$. Because of
that the first of the mentioned connections is given by the relation
\begin{equation}
\label{eq38}  n_{e;ion} (r) = \alpha Z_{i} \cdot n_{i}^{(i)} (r),
\qquad 0 < r \le r_{s;i},
\end{equation}
which corresponds to the the way of obtaining of the ion density,
and provides the satisfaction of three conditions: that the member
$n_{e;ion} (r)$ in (\ref{eqnes2}) has to be close to $\alpha Z_{i}
\cdot n_{i}^{(i)} (r)$ at least in one part of the region
$r<r_{s;i}$; in whole this region the behavior of $n_{e;ion} (r)$
has to reflects the behavior of $n_{i}^{(i)} (r)$; the ratio of the
mean numbers of additional electrons and ions in the probe particle
self-sphere has to be equal to correlation coefficient, i.e.
$\alpha$ under adopted conditions. Another of the mentioned
connections is the condition
\begin{equation}\label{eq24a}
n_{e;s} (r_{s;i}) = N_{e}(1-\alpha),
\end{equation}
which provides needed continuality of $n_{e}^{(i)} (r)$ at
$r=r_{s;i}$. The physical sense of this condition is discussed in
the further text.

The procedure of determination of the member $n_{e;s} (r)$ is
described in the second part of Appendix~\ref{sec:appb1}. The first
results of this procedure are the equation (\ref{eq39}) and the
expression (\ref{eq39a}), which provide the obtaining of the
equation for direct determination of $n_{e;s} (r)$, namely
\begin{equation}
\label{eq41}  \begin{array}{l} \displaystyle{ n_{e;s}(r) -
\frac{\kappa _{0;e}^2 }{4\pi } \cdot \frac{Z_i}{r} - \kappa _{0;e}^2
\cdot \int\limits_0^r {n_{e;s}(r')\left( {\frac{1}{r'} -
\frac{1}{r}} \right)} \cdot r'^2dr' =} \\
\displaystyle{ = n_{e;s}(r_{st} ) - \frac{\kappa _{0;e}^2 }{4\pi }
\cdot \frac{Z_i}{r_{st} } - \kappa _{0;e}^2 \cdot
\int\limits_0^{r_{st} } {n_{e;s}(r')\left( {\frac{1}{r'} -
\frac{1}{r_{st} }} \right)} \cdot r'^2dr'}.
\\
\end{array}
\end{equation}
where the screening parameter $\kappa _{0;e}$ is given by
\begin{equation}
\label{eq42}  \kappa _{0;e} = \left( {\frac{4\pi
e^2}{\displaystyle{{\partial \mu _e }/{\partial N_e }}}} \right)^{{1
\over 2}}.
\end{equation}
The next results of the used procedure are the relations
(\ref{eqB5}) and (\ref{eqB6}) which determined the member
$n_{e;s}(r)$ in the form
\begin{equation}
\label{eq45}  n_{e;s} (r) = N_{e} \cdot \left[a \cdot \frac{r_{s;i}}{r} \exp{
\left( - x_{s}\frac{r}{r_{s;i}} \right)} + b \cdot
\frac{r_{s;i}}{r} \exp{ \left( x_{s}\frac{r}{r_{s;i}}
\right)} \right], \qquad 0 < r \le r_{s;i},
\end{equation}
where the coefficients $a$ and $b$ have to satisfy the condition
(\ref{eqB7}), which provides that this relation really represents a
solution of the equation (\ref{eq41}), as well as the condition
(\ref{eqB8}), which provides the satisfying of the condition
(\ref{eq24a}). Under these conditions $a$ and $b$ are obtained in
the form
\begin{equation}
\label{eq46}  a = \frac{1 - \alpha - \frac{1}{3}
x_{s}^{2}\exp (x_{s} )}{\exp ( - x_{s} ) - \exp (x_{s} )}, \qquad
b = - \frac{1 - \alpha - \frac{1}{3} x_{s}^{2}\exp (-x_{s}
)}{\exp ( - x_{s} ) - \exp (x_{s} )},
\end{equation}
\begin{equation}
\label{eq48}  x_{s} = \kappa _{0;e} r_{s;i}.
\end{equation}
Finally, the last free parameter, i.e. the correlation coefficient
$\alpha$, is determined from the condition (\ref{eq24}) and is given
by the expression
\begin{equation}
\label{eq49}  \alpha = 1 - \frac{\textstyle{2 \over 3}x_{s}^3
}{\left( {1 + x_{s} } \right)\exp \left( { - x_{s} } \right) -
\left( {1 - x_{s} } \right)\exp \left( {x_{s} } \right)}.
\end{equation}
From here it follows that $\alpha$ satisfies the conditions
\begin{equation}
\label{eqalp}  0 <\alpha < 1, \qquad 0 < x_{s} < \infty; \qquad
\mathop {\lim}\limits_{x_{s} \to 0} \alpha  = 0; \qquad
\lim\limits_{x_{s} \to \infty }\alpha = 1;
\end{equation}
which make possible the treatment of $\alpha$ as the correlation
coefficient in the whole region $0 < x_{s} < \infty$. Finally, by
means of Eqs.~(\ref{eq22}), (\ref{eqnes2}), (\ref{eq38}) and
(\ref{eq24a}) the electron density $n_e^{(i)} (r)$ can be presented
in the form
\begin{equation}
\label{eq52}  n_e^{(i)} (r) = \left\{ {{\begin{array}{*{20}c}
 \displaystyle{ {N_e(1-\alpha) + \alpha Z_{i} \cdot n_{i}^{(i)}(r),}} \hfill & {r_{s;i} < r < \infty ,} \hfill \\
 {n_{e;s}(r)+ \alpha Z_{i} \cdot n_{i}^{(i)}(r),} \hfill & {0 < r \le r_{s;i} ,} \hfill \\
\end{array} }} \right.
\end{equation}
where $n_{i}^{(i)}(r)$ and $n_{e;s}(r)$ are given by
Eqs.~(\ref{eq34}) and (\ref{eq45})-(\ref{eq48}), which provide
$n_e^{(i)} (r)$ satisfies the condition Eq.~(\ref{eq14}), while
$a_{i}$ and $b_{i}$ are given by (\ref{eq46}) and $\alpha$- by
(\ref{eq49}).

It can be shown that $n_{e;s}(r)$, given by
(\ref{eq45})-(\ref{eq49}), in the region $r \le r_{s}$ monotonically
increases with decreasing of $r$ and satisfies the equality
\begin{equation}\label{eqdnes}
 \left[\frac{dn_{e;s}(r)}{dr}\right]_{r=r_{s;i}}=0, \qquad 0 < x_{s} <
 \infty.
\end{equation}
Apart of that, these facts provide smoothness of the electron
density $n_{e}^{(i)}(r)$ in the point $r=r_{s;i}$ and give the clear
physical sense to the condition (\ref{eq24a}) within the developed
method. Namely, we have that just the value $n_{e;s}(r_{s;i})$ in
this condition determines the value of the effective density of the
non-correlated part of the electron component, i.e. $N_{e} \cdot (1
- \alpha)$, inside and outside of the probe particle self-sphere.

\begin{figure}[htbp]
\centerline{\includegraphics[width=\columnwidth,
height=0.75\columnwidth]{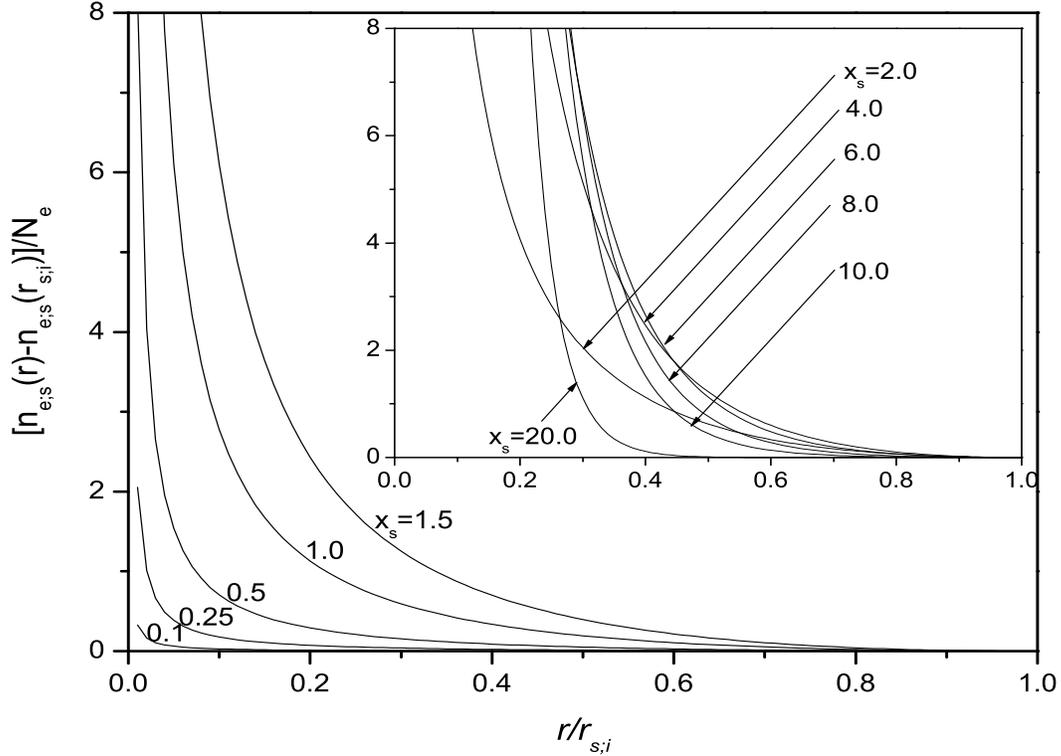}} \caption{The behavior of the
ratio $[n_{e;s}(r)-n_{e;s}(r_{s;i})]/ N_{e}$, for $0.1 \le
x_{s}=\kappa_{0;e} r_{s;i} \le 20$.} \label{fig:necor}
\end{figure}

Finally, the property (\ref{eqdnes}) of $n_{e;s}(r)$ causes such a
behavior of the difference $[n_{e;s}(r) - n_{e;s}(r_{s;i})]$, which
is illustrated by Fig.~\ref{fig:necor}. This figure shows that the
deviation of that difference from zero can be practically neglected
within the layer $0.75 r_{s;i} \le r \le r_{s;i}$ for any $x_{s} >
0$. This provides the extrapolation of the $n_{i}^{(i)} (r)$ from
the region $r > r_{s;i}$ in the part of region $r < r_{s;i}$ which
makes about $60\%$ of the probe particle self-sphere can be
performed in the same way as in Part 1, which means that the
deviation of $[n_{e;s}(r) - n_{e;s}(r_{s;i})]$ from zero can be
neglected. It is clear that such a behavior of $n_{e;sc}(r)$, which
could not be expected in advance, additionally justifies the applied
procedure.

\subsection{The case (e)}
\label{sec:mdme}

{\bf The solution $n_{e}^{(e)}(r)$.} In this case treatment of the
light (electron) and hard (ion) components appears as the main
problem. The way of its solving and the procedure of the expressing
$n_{i}^{(e)}(r)$ through $n_{e}^{(e)}(r)$ in the region $r_{s;e} < r
< \infty$ are described in first part of Appendix~\ref{sec:appbe0}.
The results of this procedure is the relation (\ref{eq66}) which can
be presented in the form
\begin{equation}
\label{eq72}  n_{i}^{(e)} (r) = N_{i}(1-\alpha) + \frac{\alpha }{Z_i
} \cdot n_{e}^{(e)} (r), \quad r_{s;e} < r < \infty.
\end{equation}
This relation, together with Eq.~(\ref{eq2}), makes possible to
express the charge density $\rho^{(e)}(r)$ in the form:
$\rho^{(e)}(r) = -e \cdot [n_{e}^{(e)}(r) - N_{e}] \cdot (1 -
\alpha)$, and repeat word-for-word the procedure from
Section~\ref{sec:mdmi}. As result we obtain the needed expression
for the electron density $n_{e}^{(e)} (r)$ in the whole space,
namely
\begin{equation}
\label{eq69}  n_{e}^{(e)} (r) = \left\{ {{\begin{array}{*{20}c}
\displaystyle{ {N_{e} - N_{e} r_{0;e} \cdot \exp (\kappa_{e} r_{0;e}
) \cdot \frac{\exp ( - \kappa_{e}
r)}{r},}} \hfill & {r_{0;e} < r < \infty ,} \hfill \\
{0,} \hfill & {0 < r \le r_{0;e} ,} \hfill \\
\end{array} }} \right.
\end{equation}
\begin{equation}
\label{eq68}  \kappa _{e} \equiv \frac{1}{r_{\kappa ;e} } = \kappa
_{0;e} \cdot \left( {1 - \alpha } \right)^{1 \over 2},
\end{equation}
where $\kappa _{0;e}$ is defined by (\ref{eq42}), and the electron
screening constant $\kappa _{e}$ and the corresponding
characteristic length $r_{\kappa;e}$ depend on the parameter $(1 -
\alpha)^{1/2}$. The radius $r_{0;e}$, similarly to the case (i), can
be presented in two equivalent forms
\begin{equation}
\label{eq70}  r_{0;e} = \gamma_{s}(x_{e}) \cdot r_{s;e}, \qquad
r_{0;e} = \gamma_{\kappa}(x_{e}) \cdot r_{\kappa;e}
\end{equation}
\begin{equation}
\label{eq71}  x_{e} = \kappa_{e} r_{s;e},
\end{equation}
where $\gamma_{s}(x)$ and $\gamma_{\kappa}(x)$ are given by
relations (28) from Part 1, and $r_{s;e}$ - by Eq.~(\ref{eqrrs}).

{\bf The solution $n_{i}^{(e)}(r)$.} Repeating the procedure from
Section~\ref{sec:mdmi} we will find the ion density $n_{i}^{(e)}(r)$
inside the probe particle self-sphere, $0 < r \le r_{s;e}$, in the
form
\begin{equation}\label{eq72a}
 n_{i}^{(e)} (r) = n_{i;s}(r) + n_{i;el}(r), \quad n_{i;el}(r)=\frac{\alpha}{Z_{i}}\cdot
n_{e}^{(e)}(r)
\end{equation}
where the member $n_{i;s}(r)$ satisfies the conditions
\begin{equation}\label{eq72b}
 \int\limits_{0}^{r_{s;e} } {n_{i;s} (r)} \cdot 4\pi r^{2}dr =
\frac{1}{Z_{i}},
\end{equation}
\begin{equation}
\label{eq75}  n_{i;s} (r_{s;e} ) = N_{i}(1-\alpha),
\end{equation}
which play similar role as the conditions (\ref{eq24}) and
(\ref{eq24a}) in the case (i).

The way of determination of $n_{i;s} (r)$ is described in second
part od Appendix~\ref{sec:appbe0}, and as the result the relations
(\ref{eqequiv0}) and (\ref{eqkappa}) are obtained. After
determination of the coefficients in the superposition
(\ref{eqequiv0}) from the conditions (\ref{eq72b}) and (\ref{eq75}),
the member $n_{i;s}(r)$ can be presented in the form
\begin{equation}\label{eq45e}
 n_{i;s} (r) = N_{i} \cdot \left[ a \cdot \frac{r_{s;e}}{r}  \exp{ \left( -
x_{s}\frac{r}{r_{s;e}} \right)} + b \cdot \frac{r_{s;e}}{r}
\exp{ \left( x_{s}\frac{r}{r_{s;e}} \right)} \right], \qquad 0 < r \le
r_{s;e},
\end{equation}
where $a$ and $b$ are given by Eq.~(\ref{eq46} and $x_{s}$ - by
Eq.~(\ref{eq48}). Finally, from Eqs.~(\ref{eq72}), (\ref{eq72a}) and
(\ref{eq45e}) it follows the expression for $n_i^{(e)} (r)$ in the
hole space, namely
\begin{equation}
\label{eq76}  n_i^{(e)} (r) = \left\{ {{\begin{array}{*{20}c}
 \displaystyle{ {N_{i}(1-\alpha) + \frac{ \alpha} {Z_{i}} \cdot n_{e}^{(e)}(r),}}
 \hfill & {r_{s;e} < r < \infty ,} \hfill \\
 \displaystyle{{n_{i;s}(r)+ \frac{ \alpha} {Z_{i}} \cdot n_{e}^{(e)}(r),}} \hfill & {0 < r \le r_{s;e} ,} \hfill \\
\end{array} }} \right.
\end{equation}
where $n_{e}^{(e)}(r)$ and $n_{i;s}(r)$ are given by
Eqs.~(\ref{eq69})-(\ref{eq71}) and (\ref{eq45e}), which provide that
$n_i^{(e)} (r)$ satisfies the condition (\ref{eq14}), and $\alpha$
is given by Eq.~(\ref{eq49}).

It can be shown that the member $n_{i;s}(r)$ monotonously increases
in the region $r < r_{s;e}$ with the decreasing of $r$ and satisfies
the equality
\begin{equation}\label{eqdnes1}
 \left[\frac{dn_{i;s}(r)}{dr}\right]_{r=r_{s;e}}=0, \qquad 0 < x_{s} < \infty,
\end{equation}
which provides smoothness of $n_{i}^{(e)}(r)$ in the point $r =
r_{s;e}$. Consequently, it guarantees that the ion density
$n_{i}^{(e)}(r)$ in the case (e) has the similar properties as the
electron density $n_{e}^{(i)}(r)$ in the case (i).

\section{The solutions $\rho^{(i;e)}(r)$ and $\Phi^{(i;e)}(r)$, and
the probe particles potential energies $U^{(i,e)}$} \label{sec:mdie}

By means Eqs.~(\ref{eq2}), (\ref{eq34}), (\ref{eq52}), (\ref{eq69})
and (\ref{eq76}) the charge densities $\rho^{(i,e)}(r)$ can be
presented in the form
\begin{equation}
\label{eq82}  \rho ^{(i,e)}(r) = \rho _{\alpha}^{(i,e)} (r) + \rho
_{\alpha;s}^{(i,e)} (r),
\end{equation}
\begin{equation}
\label{eq83}  \rho_{\alpha}^{(i,e)} (r) = Z_{e,i} e (1 - \alpha)
\cdot \left [ N_{i,e} - n_{i,e}^{(i,e)}(r)\right], \qquad 0 < r < \infty,
\end{equation}
\begin{equation}
\label{eq84}  \rho_{\alpha;s}^{(i,e)} (r) = Z_{e,i} e  \cdot
\left [ n_{e,i;s}^{(i,e)}(r) - N_{e,i} (1 - \alpha)\right], \qquad
0 < r < r_{s;i,e},
\end{equation}
where $n_{i,e}^{(i,e)}(r)$ and $n_{e,i;s}^{(i,e)}(r)$ are given by
Eqs.~(\ref{eq34}) and (\ref{eq69}), and Eqs.~(\ref{eq45}) and
(\ref{eq45e}), respectively. One can see that the expression
(\ref{eq83}) has the same structure as the corresponding expression
from Part 1 and becomes identical to it for $\alpha=0$, while
(\ref{eq84}) describes the contribution of the quantity
$[n_{e,i;s}^{(i,e)}(r) - N_{e,i}(1- \alpha)]$, which characterizes
just two-component systems.

In accordance with the structure of the expressions for
$\rho^{(i,e)}(r)$ we will find the electrostatic potentials $\Phi
^{(i,e)}(r)$, as well as the potentials $\varphi^{(i,e)}$ defined by
Eqs.~(\ref{eq7}), in the form
\begin{equation}\label{eq57p}
 \Phi ^{(i,e)}(r) = \Phi _{\alpha} ^{(i,e)}(r) + \Phi
_{\alpha;s}^{(i,e)}(r), \qquad \varphi^{(i,e)} = \varphi _\alpha
^{(i,e)} + \varphi _{\alpha;s}^{(i,e)},
\end{equation}
where the first and second members describe the contributions of the
members $\rho_{\alpha}^{(i,e)}(r)$ and $\rho_{\alpha;s}^{(i,e)}(r)$
in Eq.~(\ref{eq82}). It can be shown that by means Eqs.~(\ref{eq83})
and (\ref{eq84}), as well as Eqs.~(51)-(53) from Part 1, the members
$\Phi _{\alpha} ^{(i,e)}(r)$ and $\Phi _{\alpha;s}^{(i,e)}(r)$ can
be presented in the form

\begin{equation}
\label{eq63} \Phi_{\alpha} ^{(i,e)}(r) = \frac{Z_{i,e} e \cdot (1 -
\alpha ) }{r} \cdot \left\{ {{\begin{array}{*{20}c}
 \displaystyle{ \chi (x_{i,e} ) \cdot \exp ( { - \kappa _{i,e} r} ),} \hfill & {r_{0;i,e} < r < \infty ,} \hfill \\
 \displaystyle{{1 + \frac{ \varphi _\alpha ^{(i,e)} r }{ Z_{i,e} e \cdot (1 - \alpha
) } + \frac{1}{2} \cdot \left (  \frac{ r }{ r_{s;i,e} } \right
)^{3},}} \hfill & {0 < r \le r_{0;i,e} ,} \hfill \\
\end{array} }} \right.
\end{equation}
\begin{equation}
\label{eq63a}  \begin{array}{l} \displaystyle{\Phi_{\alpha;s}
^{(i,e)}(r) = \frac{ Z_{i,e} e }{r} \cdot \left [ \alpha + \frac{
\varphi _{\alpha;s} ^{(i,e)} r }{ Z_{i,e} e  } + 3 \cdot \frac{ a
\exp(-x_{s}\frac{r}{r_{s;i,e}}) + b
\exp(x_{s}\frac{r}{r_{s;i,e}}) + (a - b) x_{s}\frac{r}{r_{s;i,e}}}{ x_{s}^{2} }-1  \right ] - }\\
\displaystyle{ - \frac{ Z_{i,e} e }{r} \cdot \frac{ 1 - \alpha }{2}
\left( \frac{r}{r_{s;i,e}} \right )^{3} }, \qquad 0 < r \le
r_{s;i,e},
\\
\end{array}
\end{equation}
where $x_{i} $ and $x_{e} $ are defined by Eqs.~(\ref{eq36}) and
(\ref{eq71}), $x_{s}$ - by  Eq.~(\ref{eq48}), the function $\chi
(x)$ - by Eq.~(35) from Part 1. The quantities $\varphi_\alpha
^{(i,e)}$ and $\varphi _{\alpha;s}^{(i,e)}$ are obtained  by means
Eqs.~(\ref{eq83}) and (\ref{eq84}), as well as Eq.~(52) from Part 1,
and presented in the form
\begin{equation}\label{eq57a}
\varphi_{\alpha}^{(i,e)} = -Z_{i,e}e \cdot (1-\alpha) \cdot
\frac{3r_{0;i,e}}{2r_{s;i,e}^{3}} \left(r_{0;i,e}+\frac{2}{\kappa}
\right),
\end{equation}
\begin{equation} \label{eq57b}
\varphi_{\alpha ;s}^{(i,e)} = - \frac{3Z_{i,e} e}{2r_{s;i,e}} \cdot
\left[ 2 \cdot \frac{a \left( {1 - e^{ - x_{s} }} \right) + b \left(
{e^{x_{s} } - 1} \right)}{x_{s} }-(1 - \alpha )\right],
\end{equation}
where the coefficients $a$ and $b$ are given by Eqs.~(\ref{eq46}),
and the radii $r_{0;i,e}$ - by Eqs.~(\ref{eq35}), (\ref{eq36}),
(\ref{eq70}) and (\ref{eq71}), as well as Eq.~(28) from Part 1.

In accordance with Eqs.~(\ref{eq8}), (\ref{eq57p}), (\ref{eq57a})
and (\ref{eq57b}) it is appropriate to find the potential energies
$U^{(i,e)}$ of the probe particles in the cases (i) and (e) in the
form
\begin{equation}\label{eq57}
U^{(i,e)} = U_{\alpha}^{(i,e)} + U_{\alpha;s}^{(i,e)} \equiv
Z_{i,e}e \cdot \varphi_{\alpha}^{(i,e)} + Z_{i,e}e \cdot
\varphi_{\alpha;s}^{(i,e)},
\end{equation}
where the quantities $\varphi_\alpha ^{(i,e)}$ and $\varphi
_{\alpha;s}^{(i,e)}$ are given by Eqs.~(\ref{eq57a}), (\ref{eq57b}).
It can be shown that, because of the structure of Eq.~(\ref{eq57a}),
the member $U_{\alpha}^{(i,e)}$ can be taken as:
$U_{\alpha}^{(i,e)}=(1-\alpha)\cdot U$, with $U$ given by one of two
equivalent expressions (32) and (33) from Part 1, where the
parameters $Z$, $x$, $r_{\kappa}$ and $r_{s}$ are replaced by
$Z_{i,e}$, $x_{i,e}$, $r_{\kappa;i,e}$ and $r_{s;i,e}$. The figure
~(\ref{fig:U}) illustrates the behavior of the ratio of the probe
particle potential energy, determined by means to
Eqs.~(\ref{eq57a}), (\ref{eq57b}) and (\ref{eq57}), and the
corresponding DH energy, defined by Eq.~(\ref{eqA6}), in the case of
the classical plasma with $Z_{i} = 1$ and $ T_{e} = T_{i}$. This
figure shows that in two-component case the DH region, i.e. the
region where this ratio is close to unity, principally does not
exist.

\section{Results and discussions}
\label{sec:pso}

\begin{figure}[htbp]
\centerline{\includegraphics[width=\columnwidth,
height=0.75\columnwidth]{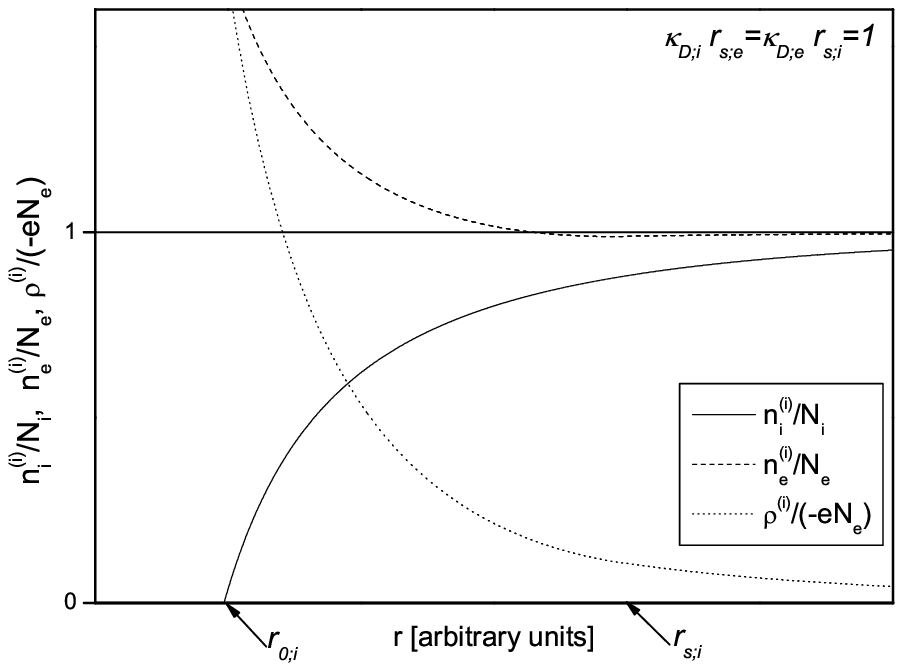}} \caption{The behavior of
reduced densities $n_{i}^{(i)} (r) / N_{i}$, $n_{e}^{(i)} (r) /
N_{e}$ and $\rho^{(i)} (r) / (-eN_{e}) $ in the case $Z_{i}=1$,
$T_{e}=T_{i}$ and $\kappa_{0;i} r_{s;i} = 1$, where $\kappa_{0;i} $
is defined by (\ref{eq29}).} \label{fig:nrho}
\end{figure}

The expressions (\ref{eq34})-(\ref{eq52}) and
(\ref{eq69})-(\ref{eq63}) show that:{\it the obtained solutions
satisfy all conditions from Section~\ref{sec:tas}, are free of all
non-physical properties of DH solutions, and possesses the positive
properties (b1) and (b2), noted in Section~1 of Part 1}. The
behavior of electron, ion and the charge density is illustrated by
Fig.~\ref{fig:nrho}. The way of their obtaining provides the
possibility of their application, independently of the treatment
(classical or quantum mechanical) of electron and ion components of
the considered systems. Apart of that, the structure of these
expressions makes possible their application not only to
electron-ion plasmas, but also to ion-ion two-component systems
(e.g. some electrolytes).
\begin{figure}[htbp]
\centerline{\includegraphics[width=\columnwidth,
height=0.75\columnwidth]{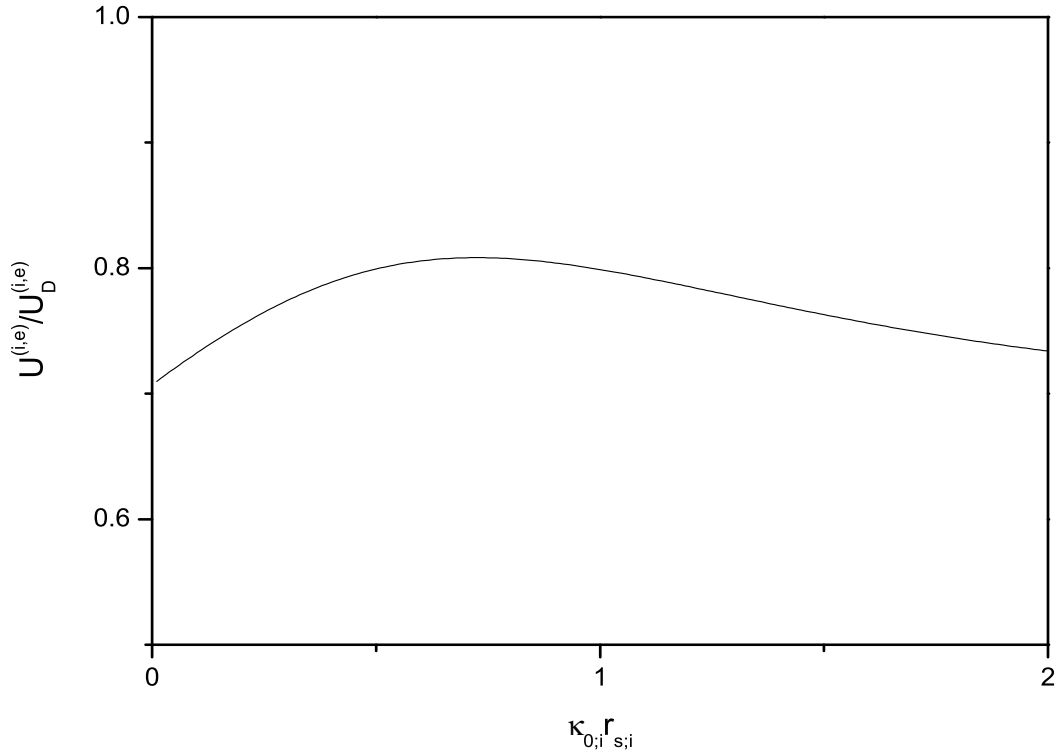}} \caption{The behavior of the
ratio $U^{(i,e)}/U_{D}^{(i,e)}$ as a function of the parameter
$\kappa_{0;i}r_{s;i}$ in the case of completely classical plasma
with $Z_{i}=1$ and $T_{i}=T_{e}$, where $\kappa_{0;i}$ is defined by
Eq.~(\ref{eq29}).} \label{fig:U}
\end{figure}

Since Eqs.~(\ref{eq45}), (\ref{eq52}), (\ref{eq45e}) and
(\ref{eq76}) show that the solutions $n_{e,i}^{(i,e)}(r)$ are
singular in the point $r = 0$, it is useful to note that the
existence of singularities in model solutions is fully acceptable,
if it has not other non-physical consequences. Such solutions are
well known in physics: it is enough to mention, for example,
Thomas-Fermi's models of electron shells of heavy atoms
(\cite{tho26,fer28}; see also \cite{gom50}), which use in plasma
research till now (see e.g. \cite{men02}).

In the case of two-component system are obtained the parameters
$r_{0;i,e}$, $\gamma_{s}(x_{i,e})$ and $\gamma_{\kappa}(x_{i,e})$,
given by Eqs.~(\ref{eq35}), (\ref{eq36}), (\ref{eq70}) and
(\ref{eq71}), analogous to that ones from Part 1. In accordance with
Eq.~(28) from Part 1 the parameters $r_{0;i}$ and $r_{0;e}$ satisfy
the conditions
\begin{equation}\label{eqnon1}
 0 < r_{0;i,e} < r_{s;i,e}, \quad 0 < x_{i;e} < \infty; \qquad \mathop {\lim
}\limits_{x_{i,e} \to 0} r_{0;i,e}  = 0; \qquad \lim\limits_{x_{i,e} \to
\infty }r_{0;i,e} = r_{s;i,e};
\end{equation}
where $r_{s;i,e}$ is given by Eq.~(\ref{eqrrs}), and $x_{i}$ and
$x_{e}$ - by Eqs.~(\ref{eq36}) and (\ref{eq71}). These conditions
make possible the treatment of $r_{0;i}$ and $r_{0;e}$ as the radii
of the spheres centered on the probe particles, which are
classically forbidden for the free charged particles from their
neighborhoods, and $\gamma_{s}(x_{i,e})$ and
$\gamma_{\kappa}(x_{i,e})$ - some kind of non-ideality parameters
(see Part 1). From (\ref{eqnon1}) it follows also that
Eqs.~(\ref{eq34}), (\ref{eq52}), (\ref{eq69}) and (\ref{eq63}) for
electron and ion densities are {\it applicable to the two-component
systems with any non-ideality degree}.

Also, in this paper is obtained the quantity $\alpha(x_{s})$,
defined by Eqs.~(\ref{eq48}) and (\ref{eq49}), which has the sense
of the coefficient of electron-ion correlation. Let us note that two
simple approximative expressions for $\alpha(x_{s})$, which serve
very well in wide region of $x_{s}$, are given in Part~3.

One of the most important results of this papers is establishing of
the fact that in two-component plasmas ion and electron components
have to be described exceptionally by means of screening constants
$\kappa_{i}$ and $\kappa_{e}$, and the corresponding screening radii
$r_{\kappa;i}$ and $r_{\kappa;e}$ which are introduced in this
paper. This means that {\it Debye-H\"{u}ckel's screening constant}
$\kappa_{D}$ {\it and the radius} $r_{D}$ {\it do not appear in the
theory and, consequently, principally do not have the physical
sense}. It is confirmed by Fig.~\ref{fig:U} where the probe particle
potential energies obtained here are compared with the corresponding
DH values. Finally, comparing of the expressions (\ref{eq57p}),
(\ref{eq63}) and (\ref{eq63a}) for the potentials $\Phi^{(i,e)}(r)$
with the expression (\ref{eqA3}) for DH solutions
$\Phi_{D}^{(i,e)}(r)$ shows that in two-component case the principal
difference between these solutions there is not only inside the
probe particle self-spheres, but also in the rest of space. Namely,
out of these self-spheres $\Phi_{D}^{(i,e)}(r) \sim
\exp(-\kappa_{D})r$, where $\kappa_{D}$ is given by
Eq.~(\ref{eqA1}), while the solution $\Phi^{(i,e)}(r) \sim
\exp(-\kappa_{i,e})r$, where $\kappa_{i}$ and $\kappa_{e}$ are given
by Eq.~(\ref{eq29}), (\ref{eq42}) and (\ref{eq68}). This fact
justifies the usage in \cite{dra65,bow61,vit90,vit01} the constants
which are close to $\kappa_{i}$ and $\kappa_{e}$ instead of DH
constants.

\begin{acknowledgments}
The authors are thankful to the University P. et M. Curie of Paris
(France) for financial support, as well as to the Ministry of
Science of the Republic of Serbia for support within the Project
141033 "Non-ideal laboratorial and ionospheric plasmas: properties
and applications".
\end{acknowledgments}

\begin{appendix}

\section{DH solutions}
\label{sec:appa}

In DH method it is assumed that for determination of $n_{e}^{(i)}$
and $n_{i}^{(e)}$, apart of Eqs.~(\ref{eq9}), it can be applied the
equations
\begin{equation}
\begin{array}{c}
\label{eq10}  n_{e,i}^{(i,e)} (r) - N_{e,i} = {\displaystyle -
\frac{Z_{e,i} e}{\partial \mu _{e,i}/\partial N_{e,i}}\Phi
^{(i,e)}(r)}, \qquad {\displaystyle {\partial \mu_{e,i}}/{\partial
N_{e,i}} \equiv \left [\frac{\partial \mu_{e,i}
(n,T_{e,i})}{\partial n}\right]_{n = N_{e,i}}},
\end{array}
\end{equation}
which represent the linearized form of the equations: $\mu_{e,i}
\left( {n_{e,i}^{(i,e)} (r),T_{e,i} } \right) + Z_{e,i} e \cdot
\Phi^{(i,e)} (r)=\mu _{e,i} \left( {N_{e,i} ,T_{e,i} } \right)$.
Within DH method equations (\ref{eq9}) and (\ref{eq10}) apply
together in order to express the charge densities in the Poison's
equations (\ref{eq5}) through the electrostatic potential in the
whole region $0 < r < \infty$. That way, in accordance with
Eqs.~(\ref{eq1})-(\ref{eq3}) and (\ref{eq5}) one obtains Helmholtz's
equation
\begin{equation}\label{eqA1}
\nabla^{2} \Phi ^{(i,e)}(r) = \kappa _D^2 \Phi ^{(i,e)}(r), \qquad
\kappa _D \equiv \frac{1}{r_{D}} =\left( {\kappa _{0;i}^2 + \kappa
_{0;e}^2 } \right)^{\frac{1}{2}},
\end{equation}
where $\kappa_{D} $ is DH screening constant, and $\kappa _{0;i}$
and $\kappa _{0;e} $ are the partial screening constants defined by
Eqs.~(\ref{eq29}) and (\ref{eq42}).

DH electrostatic potential $\Phi _D^{(i,e)} (r)$ represent the
solutions of equation (\ref{eqA1}), which is obtained in the whole
space under the boundary conditions (\ref{eq6}) and (\ref{eq7}).
Then, by means of Eqs.~(\ref{eq9}) and (\ref{eq10}) DH solutions for
the charge and particle densities are obtained. All these solutions
are given by the expressions
\begin{equation}\label{eqA3}
 \Phi _D^{(i,e)} (r) = \frac{Z_{i,e} e}{r}\exp ( - \kappa _{D}
r), \qquad \rho _D^{(i,e)} (r) = - \frac{Z_{i,e} e\kappa _{D}^2
}{4\pi }\frac{\exp ( - \kappa _{D} r)}{r},
\end{equation}
\begin{equation}\label{eqA4}
 n_{D;i}^{(i)} (r) = N_i - \frac{\kappa _{0;i}^2 }{4\pi
}\frac{\exp ( - \kappa _{D} r)}{r}, \qquad n_{D;e}^{(e)} (r) = N_e -
\frac{\kappa _{0;e}^2 }{4\pi }\frac{\exp ( - \kappa _{D} r)}{r},
\end{equation}
\begin{equation}\label{eqA5}
 n_{D;e}^{(i)} (r) = N_e + \frac{Z_{i}\kappa _{0;e}^2 }{4\pi
}\frac{\exp ( - \kappa _{D} r)}{r}, \qquad n_{D;i}^{(e)} (r) = N_i +
\frac{\kappa _{0;i}^2 }{4\pi Z_{i}}\frac{\exp(- \kappa _{D} r)}{r},
\end{equation}
where the screening constant $\kappa _{D} $ is given by
(\ref{eqA1}). From Eqs.~(\ref{eq7}), (\ref{eq8}) and (\ref{eqA3}) it
follows that DH potential $ \varphi _D^{(i,e)} = - Z_{i,e} e \cdot
\kappa _{D}$ and consequently, DH potential energy $U_D^{(i,e)}$ is
given by
\begin{equation}\label{eqA6}
U_D^{(i,e)} = - Z_{i,e} e \cdot \kappa_{D}= - \frac{(Z_{i,e}
e)^2}{r_{D}},
\end{equation}
where $\kappa_{D}$ and $r_{D}$ are defined by Eq.~(\ref{eqA1}).

\section{The system $S_{a}^{(i)}$}
\label{sec:appb1}

{\bf The region $r_{s;i} < r < \infty$.} We will start from the fact
that the different electron-ion and dusty plasmas can be
successfully described in the approximation of fixed heavy charged
particles (see e.g. \cite{hua63,dju91,ada94,mai00,gor02}). In
accordance with this, we will treat the electronic component of the
system $S_{a}^{(i)}$ considering all ions as immobile with respect
to electrons, keeping in mind that in such a case the difference
between the probe particle and ions does not exist regarding the
electrons. In Fig.~\ref{fig:scheme} the several of ions self-spheres
(sphere with the radius $r_{s;i}$, centered on ions) in the
neighborhood of the probe particle (the point $O$) are schematically
shown. This figure should make more apparent the fact that the
behavior of the electron inside the self-sphere of an ion in the
point $O'$, far from the point $O$, has to be almost completely
caused by its interaction with few ions which are close to the point
$O'$. Because of that, in further considerations all ions are
treated equally with the probe particle.

Here, we will denote by ($^{i;\ast}$) the possible ion configuration
in the system $S_{a}^{(i)}$, and by $n_{e}^{(i;\ast)} (\vec {r})$
and $n_{i}^{(i;\ast)}(\vec {r})$- the corresponding electron and ion
densities. Also, we will denote by $\bar n_{e;out}^{(i; \ast)}$ an
average value of $n_{e}^{(i;\ast)} (\vec {r})$ within the part of
space consisting of the outing of probe particle's and all ion's
self-spheres (see shadowed area in figure \ref{fig:scheme}). The
condition of the thermodynamical equilibrium of the electron
component in the case of a configuration ($^{i;\ast}$) has the form
\begin{equation}\label{eqmub1}
 \mu_{e} \left( {n_{e}^{(i;\ast )} (\vec {r}),T_e } \right) + ( -
e) \cdot \Phi^{(i;\ast )} (\vec {r}) = \mu_{e} \left(
{n_{e}^{(i;\ast )} (\vec {r}_{st}^{i;\ast} ),T_e } \right) + ( - e)
\cdot \Phi ^{(i;\ast )} (\vec {r}_{st}^{i;\ast} ),
\end{equation}
where $\Phi^{(i;\ast )} (\vec {r})$ is the corresponding total
electrostatic potential, and $\vec {r}_{st}^{(i;\ast)}$ is any point
where $n_{e}^{(i;\ast )} (\vec {r}) = \bar n_{e;out}^{(i;*)}$.
Finally, taking the chemical potentials $\mu_{e} ( {n_{e}^{(i;\ast
)} (\vec {r}),T_e} )$ in (\ref{eqmub1}) as series of the differences
$[ n_{e}^{(i;\ast )} (\vec {r}) - N_{e}]$ along with keeping only
first two members, we obtain $n_{e}^{(i;\ast )}(\vec{r})$ in the
form
\begin{equation}\label{eqmub2}
 n_{e}^{(i;\ast )}(\vec{r}) = \bar n_{e;out}^{(i;\ast )} + \frac
{-e}{(\partial \mu_{e}/ \partial N_{e})} \cdot \left[ \Phi ^{(i;\ast
)}(\vec {r_{st}}^{(i; \ast)}) - \Phi ^{(i;\ast )}(\vec{r}) \right].
\end{equation}
In contrast to (\ref{eq9}), the equation (\ref{eqmub2}) can be
applied practically in the whole space, which is allowed by the
behavior of the electron component in the presences of the positive
charged particles.

In accordance with the described picture, the sought electron and
ion densities $n_{e}^{(i)}(r)$ and $n_{i}^{(i)}(r)$ are treated here
as a result of averaging of all densities $n_{e}^{(i;\ast )} (\vec
{r})$ and $n_{i}^{(i;\ast)}$. Because of the structure of
Eq.~(\ref{eqmub2}) we will have that: $ n_{e}^{(i)} (r) = N_{e;st} +
n_{e;d} (r)$, where $N_{e;st}$ and $n_{e;d} (r)$ are the mean values
of the first and second members in Eq.~(\ref{eqmub2}). Taking the
member $n_{e;d} (r)$ as series of the differences $[ n_{i}^{(i)} (r)
- N_{i}]$ along with keeping only first two members, we obtain that:
$n_{e}^{(i)} (r) = N_{e;st} + n_{e;d} (r=\infty) + K \cdot [
n_{i}^{(i)} (r) - N_{i}]$, where accordingly to the boundary
condition (\ref{eq3}) the member $n_{e;d} (r=\infty)$ has to satisfy
the equality $n_{e;d} (r=\infty)= N_{e} - N_{e;st}$. Finally, taking
that $K=\alpha \cdot Z_{i}$ we obtain the mean electron density
$n_{e}^{(i)} (r)$ in the form
\begin{equation}
\label{eq:22d}  n_{e}^{(i)} (r) = N_{e} +
   \alpha \cdot Z_{i} [ {n_{i}^{(i)} (r) - N_{i} } ].
\end{equation}
This relation shows that $n_{e}^{(i)} (r)$ in the region $r
> r_{s;i}$ can be expressed by means only one unknown parameter
$\alpha$.

{\bf The region $0 < r < r_{s;i}$: the basic equation.} Let
($^{s;\ast}$) denotes such an ion configuration in which the probe
particle self-sphere is left free of ions, and $n_{e}^{(s;\ast)}
(\vec {r})$ - the corresponding electron density. The member
$n_{e;s}(r)$ in (\ref{eqnes2}) will be identified with the result of
averaging of densities $n_{e}^{(s;\ast )} (\vec {r})$ over all
configurations ($^{s;\ast}$) in the region $0 < r \le r_{s;i}$.

The corresponding condition of the thermodynamical equilibrium in
the case of a configuration ($^{s;\ast}$) obtains from
(\ref{eqmub1}) replacing the index (i,*) by (s,*) and fixing the
point $\vec{r}_{st}^{(s,*)}=\vec{r}_{st}$, where $r_{st} \le
r_{s;i}$. Then, repeating the procedure of the obtaining of Eq.~
(\ref{eqmub2}), we obtain the equation
\begin{equation}
\label{eq39}  n_{e;s} (r) - n_{e;s}(r_{st} ) =  \frac{ -e} {\partial
\mu _e /\partial N_e}\cdot \left[\Phi _{s}(r_{st}) -
\Phi_{s}(r)\right].
\end{equation}
where the potentials $\Phi_{s} (r)$ and $\Phi_{s} (r_{st})$ in the
considered region are given by Eq.~(53) from Part 1 with the charge
density $(- e) \cdot n_{e;s} (r)$. Since all information about the
outing of the probe particle self-sphere ($r_{s;i} < r < \infty$)
contained in the same constant members in expressions for
$\Phi_{s}(r)$ and $\Phi _{s}(r_{st})$, the difference $\left[\Phi
_{s}(r_{st}) - \Phi_{s}(r)\right]$ in Eq.~(\ref{eq39}) is given by
\begin{equation}
\label{eq39a}
\begin{array}{l} \displaystyle{\Phi _{s}(r_{st}) - \Phi_{s}(r) =
\frac{Z_{i}e}{r_{st}} - \frac{Z_{i}e}{r} -
4\pi\int\limits_0^{r_{st}}(- e) \cdot n_{e;s}
(r^{'}\left(\frac{1}{r^{'}} - \frac{1}{r_{st}}\right) {r^{'}}^{2}
dr{'} + }\\
\displaystyle{+ 4\pi\int_{0}^{r}(- e) \cdot n_{e;s}
(r^{'}\left(\frac{1}{r^{'}} - \frac{1}{r}\right) {r^{'}}^{2} dr{'}}.\\
\end{array}
\end{equation}
From Eqs.~ (\ref{eq39}) and (\ref{eq39a}) it follows the equation
(\ref{eq41}) for direct determination of the member $n_{e;s} (r)$.
In order to find the solution of the equation (\ref{eq41}), we will
introduce the function $S(r)$ given by relation
\begin{equation}\label{eqB1}
n_{s}(r) - \frac{\kappa _{0;e}^2 }{4\pi } \cdot \frac{Z_i}{r} =
\frac{S(r)}{r}.
\end{equation}
Consequently, Eq.~(\ref{eq41}), after the multiplication by $r$,
transforms to the equation
\begin{equation}\label{eqB2}
 \begin{array}{l} \displaystyle{ S(r) - \kappa _{0;e}^2
\int\limits_0^r {S(r')} \left( {r - r'} \right)dr' - \frac{Z_i \cdot
\kappa _{0;e}^4 }{4\pi }\int\limits_0^r
{\left( {r - r'} \right)dr' = }} \\
\displaystyle{ = \frac{r}{r_{st}} \cdot \left[ S(r_{st}) - \kappa
_{0;e}^2 \int\limits_{0}^{r_{st}} S(r') \left( r_{st} - r'
\right)dr' - \frac{Z_i \cdot \kappa _{0;e}^4 }{4\pi
}\int\limits_{0}^{r_{st}}
\left( r_{st} - r' \right)dr' \right] }. \\
 \end{array}
\end{equation}
Applying the operator $\frac{d^2}{dr^2}$ to left and right sides of
{\ref{eqB2}} we obtain the equation
\begin{equation}\label{eqB3}
 \frac{d^2S(r)}{dr^2} - \kappa _{0;e}^2 S(r) - \kappa
_{0;e}^2 \cdot \frac{Z_i \kappa _{0;e}^2 }{4\pi } = 0,
\end{equation}
which can be presented in the form
\begin{equation}\label{eqB4}
 \frac{d^2}{dr^2}\left[ {S(r) + \frac{Z_i \kappa _{0;e}^2 }{4\pi
}} \right] = \kappa _{0;e}^2 \left[ {S(r) + \frac{Z_i \kappa
_{0;e}^2 }{4\pi }} \right].
\end{equation}
From Eqs.~(\ref{eqB1}) and (\ref{eqB4}) it follows that in the
general case $n_{s}(r)$ is given by the relation
\begin{equation}\label{eqB5}
 n_{s}(r) = \frac{A \cdot \exp \left( { - \kappa _{0;e} r}
\right) + B \cdot \exp \left( {\kappa _{0;e} r} \right)}{r}.
\end{equation}
Taking the coefficients $A$ and $B$ in the form
\begin{equation}\label{eqB6}
 A =  r_{s;i} \cdot N_{e} \cdot a, \qquad B = r_{s;i} \cdot N_{e} \cdot b,
\end{equation}
we will present Eq.~(\ref{eq41}) in the form
\begin{equation}\label{eqB61}
 \begin{array}{l} \displaystyle{ a \cdot \exp \left( { - \kappa
_{0;e} r} \right) + b \cdot \exp \left(
{\kappa _{0;e} r} \right) - \frac{Z_i \kappa _{0;e}^2 }{4\pi r_{s;i}N_{e}} + }\\
\displaystyle{ + \kappa _{0;e}^2 \int\limits_0^r {\left[ {a \cdot
\exp \left( { - \kappa _{0;e} r'} \right) + b \cdot \exp \left(
{\kappa _{0;e} r'} \right)}
\right]} \cdot \left( {r - r'} \right)dr' = }\\
\displaystyle{ = \frac{r}{r_{st} }\left\{ {a \cdot \exp \left( { -
\kappa _{0;e} r_{st} } \right) + b \cdot \exp \left( {\kappa _{0;e}
r_{st} } \right) - \frac{Z_i
\kappa _{0;e}^2 }{4\pi r_{s;i}N_{e}} + } \right.} \\
\displaystyle{ \left. { + \kappa _{0;e}^2 \int\limits_0^{r_{st} }
{\left[ {a \cdot \exp \left( { - \kappa _{0;e} r'} \right) + b \cdot
\exp \left( {\kappa _{0;e}
r'} \right)} \right]} \cdot \left( {r_{st} - r'} \right)dr'} \right\} .} \\
 \end{array}
\end{equation}
From here it follows the equation
\begin{equation}\label{eqB62}
 \begin{array}{l} \displaystyle{ \left( {b - a }
\right)\kappa _{0;e} r + \left( {a + b - \frac{Z_i
\kappa _{0;e}^2 }{4\pi r_{s;i}N_{e}}} \right) = }\\
\displaystyle{ \frac{r}{r_{st} }\left[ {\left( {b - a }
\right)\kappa _{0;e} r_{st} + \left( {a + b - \frac{Z_i \kappa
_{0;e}^2 }{4\pi r_{s;i}N_{e}}} \right)}
\right], }\\
 \end{array}
\end{equation}
which can be presented in the form
\begin{equation}\label{eqB63}
 a + b - \frac{Z_i \kappa _{0;e}^2 }{4\pi r_{s;i}N_{e}} =
\frac{r}{r_{st} }\left( {a + b - \frac{Z_i \kappa _{0;e}^2 }{4\pi
r_{s;i}N_{e}}} \right),
\end{equation}
where in the general case $r \ne r_{st}$. This means that the
coefficients $a$ and $b$ have to satisfy the condition
\begin{equation}\label{eqB7}
 a + b = \frac{Z_i \kappa _{0;e}^2 }{4\pi r_{s;i}\cdot N_{e}}\equiv
 \frac{(\kappa_{0;e}r_{s;i})^{2}}{3},
\end{equation}
which provides that unknown parameter $r_{st}$ disappears from
further considerations, and that $\rho_{s}(r)$ given by
Eq.~(\ref{eqB5}) and Eq.~(\ref{eqB6}) really satisfies the equation
(\ref{eq41}). Then, from Eqs.~(\ref{eqB5}) and (\ref{eqB6}) it
follows the sought expression (\ref{eq45}) for the member
$n_{e;s}(r)$. Finally, from Eqs.~(\ref{eq24a}) and (\ref{eq45}) it
is obtained the other condition which is necessary for the
determination $a$ and $b$, namely
\begin{equation}\label{eqB8}
a \cdot \exp \left( { - \kappa _{0;e} r_{s;i} } \right) + b \cdot
\exp \left( {\kappa _{0;e} r_{s;i} } \right) = 1-\alpha,
\end{equation}
where the free parameter $\alpha$ has to be determined from the
condition Eq.~(\ref{eq24}).

\section{The system $S_{a}^{(e)}$} \label{sec:appbe0}

{\bf The region: $r_{s;e} < r < \infty$.} In the case (e) we will
start from the fact that the fixed probe particle with the charge
$(-e)$ {\it principally can not be represent of a free electron in
the system} $S_{in}$ {\it from the aspect of the interaction with
positive ions}, since their average distribution in the neighborhood
of such a particle would resemble to a distribution of positive ions
in plasma in the neighborhood of a heavy negative ion.

The exit from this situation is the treatment of accessory system
$S_{a}^{(e)}$ as a single component one with the electron gas on the
corresponding positive charged background, which, contrary to the
background described in Part 1, is not homogeneous. We keep in mind
the background with the charge density taken in the form $Z_{i}e
\cdot n_{i}^{(e)}(r)$, which in the classical case is able to model
the average distribution of positive charge in the neighborhood of
an electron in the system $S_{in}$. If sought distribution is found,
the factor $n_{i}^{(e)}(r)$ could be treated as the corresponding
ion density. Here we use the fact that such a distribution can be
found since existing conditions which establish correspondence
between the systems $S_{a}^{(e)}$ and $S_{in}$ are enough for
determination of all needed parameters.

One of the mentioned conditions is that the behavior of
$n_{e}^{(e)}(r)$ and $n_{i}^{(e)}(r)$ reflect the existence of the
electron-ion correlation (discussed in Section~\ref{sec:mdmi}) which
is characterized by the coefficient $\alpha$. We will take into
account that this coefficient in the case $(e)$ can be treated as
the probability of the following event: the decreasing of number of
electrons for $Z_{i}$ in the region $r_{s;e} < r < \infty$ coincides
with the decreasing of the number of ions for unity. This means that
in this case we have the relation
\begin{equation}
\label{eq66}  \alpha =  \frac{[N_{e} - n_{e}^{(e)} (r) ]/Z_{i} }
{N_{i}  - n_{i}^{(e)} (r)},
\end{equation}
which connects $n_{e}^{(e)}(r)$ and $n_{i}^{(e)}(r)$ in the region
$r_{s;e} < r < \infty$ by means of the known parameter $\alpha$. In
Section~\ref{sec:mdme} the relation (\ref{eq66}) is taken as the
starting point for the determination $n_{i}^{(e)}(r)$ and
$n_{e}^{(e)}(r)$ outside of probe particle self-sphere in the system
$S_{a}^{(e)}$.

{\bf The region: $0 < r < r_{s;e}$.} In order to determine the
member $n_{i;s} (r)$ in Eq.~(\ref{eq72a}) we will take into account
the following facts: the form of $n_{i;s} (r)$ has to be similar to
the form of the member $n_{e;s} (r)$ in the case (i); the parameters
which characterize $n_{i;s} (r)$ have to be closely connected with
the parameters which characterize $n_{e;s} (r)$; the procedure of
obtaining of $n_{i;s} (r)$ has to provide automatic applicability
for all possible $N_{e}$, $Z_{i}$, $T_{e}$ and $T_{i}$ and
self-consistence of final expressions. In accordance with this, the
member $n_{i;s} (r)$ will be taken as a superposition of two
dimensionless functions of dimensionless argument $(r/r_{s;e})$,
given by relations
\begin{equation}\label{eqequiv0}
n_{i;s} (r) = a_{i} \cdot
(r/r_{s;e})^{-1}\exp{[-(\kappa_{s;i}r_{s;e}){r}/{r_{s;e}}]} + b_{i}
\cdot (r/r_{s;e})^{-1}\exp{[(\kappa_{s;i}r_{s;e}){r}/{r_{s;e}}]},
\end{equation}
\begin{equation}\label{eqkappa}
 \kappa_{s;i} =
\kappa_{0;e}\cdot\frac{r_{s;i}}{r_{s;e}}=\kappa_{0;e} \cdot
Z_{i}^{\frac{1}{3}},
\end{equation}
where $\kappa_{0;e}$ is given by (\ref{eq42}), and the screening
constant $\kappa_{s;i}$ is chosen in such a manner which guarantees
that the relation
\begin{equation}\label{eqequiv}
\frac{r'}{r_{s;i}}\exp{(\mp
\kappa_{s;i}r_{s;e}\cdot\frac{r'}{r_{s;i}})}=\frac{r"}{r_{s;e}}\exp{(\mp
\kappa_{0;e}r_{s;i}\cdot\frac{r"}{r_{s;e}})}
\end{equation}
is valid for any $(r'/r_{s;i})=(r"/r_{s;e})$, when $0 < r' <
r_{s;i}$ and $0 < r" < r_{s;e}$. The coefficients $a_{i}$ and
$b_{i}$ in Eq.~(\ref{eqequiv0}) have to be found from the conditions
(\ref{eq72b}) and (\ref{eq75}).

\section{The interpretation of the systems considered}
\label{sec:appc} Let $S_{in;M_{i}}$ be the model finite spherical
system with total ion and electron numbers $M_{i}$ and
$M_{e}=Z_{i}M_{i}$, where $M_{i}$ is an integer number, and with the
radius $R_{M_{i}}$ and volume $V_{M_{i}}$ determined by relation:
$M_{i}/V_{M_{i}}=N_{i}$. From here it follows that the system
$S_{in;M_{i}}$ is neutral as a whole and
$M_{e}/V_{M_{i}}=N_{e}=Z_{i}N_{i}$. We will assume that in
$S_{in;M_{i}}$ the ion and electron components can be treated as
gases in states of the thermodynamical equilibrium with temperatures
$T_{i}$ and $T_{e}$. In a usual way, we will treat the basic system
$S_{in}$ as a thermodynamical limit of the systems $S_{in;M_{i}}$ ,
i.e. as the result of transition: $M_{i} \to \infty $ and $V_{M_{i}}
\to \infty $, under conditions
\begin{equation}\label{eqC0}
 M_{i}/V_{M_{i}}= N_{i}, \qquad T_{i} = const., \qquad T_{e} =
const.
\end{equation}
In the case $(i)$, we will associate with every system
$S_{in;M_{i}}$ an other system $S_{a;M_{i}}^{(i)}$, which differs
from $S_{in;M_{i}}$ only by the change of one of the free ions for
the probe particle, with the same charge $Z_{i}e$ fixed in the
center of that system.

We will take into account that the systems $S_{a;M_{i}}^{(i)}$ are
also neutral as a whole. From here follows the relation
\begin{equation}\label{eqC1}
 Z_{i}e + Z_{i}e \cdot \int\limits_0^{R_{M_{i}}} {n_{i;M_{i}}(r)
\cdot 4\pi r^2} dr - e \cdot N_{e} V_{M_{i}} = 0,
\end{equation}
where $n_{i;M_{i}}(r)$ are the corresponding mean local ion
densities, and
\begin{equation}\label{eqC2a}
 N_{e} V_{M_{i}} = Z_{i}N_{i}\cdot \frac {4\pi}{3} R_{M_{i}}^{3}
\equiv Z_{i} \cdot \int\limits_0^{R_{M_{i}}} N_{i} \cdot 4\pi r^2
dr.
\end{equation}
From (\ref{eqC1}) and (\ref{eqC2a}), after their multiplication with
$(Z_{i}e)^{-1}$ the equation follows
\begin{equation}\label{eqC3a}
 \int\limits_0^{R_{M_{i}}} [ N_{i}- n_{i;M_{i}}(r)] \cdot 4\pi
r^2 dr = 1.
\end{equation}
Based on it, we have it that
\begin{equation}\label{eqC4a}
 \lim\limits _{M_{i} \to \infty }\int\limits_{0}^{R_{M_{i}}}
\left[ N_{i}-n_{i;M_{i}}(r)\right] \cdot 4\pi r^{2} dr =
\int\limits_{0}^{\infty} \left[ N_{i}-n_{i;\infty}(r)\right] \cdot
4\pi r^{2} dr = 1,
\end{equation}
where $n_{i;\infty}(r) = \mathop {\lim }\limits_{M_{i} \to \infty }
n_{i;M_{i}} (r)$ for any $r \ge 0$. Here, we will treat just
$n_{e;\infty} (r)$ as the mean local ion density in the accessory
systems $S_{a}^{(i)}$, i.e. as $n_{i}^{(i)}(r)$. In accordance with
this, from (\ref{eqC4a}) it directly follows that $n_{i}^{(i)}(r)$
has to satisfy the condition (\ref{eq13}) for the case $(i)$.

Since a similar reasoning may be repeated in the case $(e)$, we can
consider the condition (\ref{eq13}) must be satisfied in both $(i)$
and $(e)$ cases. Finally, it is clear that in the case when one of
components in described systems is changed with the corresponding
background, we obtain the equation which corresponds to Eq.~(21)
from Part 1.

\end{appendix}


\newcommand{\noopsort}[1]{} \newcommand{\printfirst}[2]{#1}
  \newcommand{\singleletter}[1]{#1} \newcommand{\switchargs}[2]{#2#1}

\end{document}